%% file: SST-1M-TelCTL.tex
\documentclass[hyper]{PoS}
\input{myaddontotemplates.tex}

\setboolean{AC@nolist}{true} 
\usepackage{lineno}
\newboolean{showevent}
\setboolean{showevent}{True}
\newboolean{showLEs}
\setboolean{showLEs}{False}

\title{Software design for the control system for Small-Size Telescopes with single-mirror of the Cherenkov Telescope Array}
\ShortTitle{Software design for SST-1M}
\input{authors_compact.tex}

\abstract{The Small-Size Telescope with single-mirror (SST-1M) is a 4 m Davies-Cotton telescope and is among the proposed telescope designs for the Cherenkov Telescope Array (CTA). It is conceived to provide the high-energy ($>$ few \TeV) coverage. The SST-1M contains proven technology for the telescope structure and innovative electronics and photosensors for the camera. Its design is meant to be simple, low-budget and easy-to-build industrially.
	
Each device subsystem of an SST-1M telescope is made visible to CTA through a dedicated industrial standard server. The software is being developed in collaboration with the CTA Medium-Size Telescopes to ensure compatibility and uniformity of the array control. Early operations of the SST-1M prototype will be performed with a subset of the CTA central array control system based on the Alma Common Software (ACS). The triggered event data are time stamped, formatted and finally transmitted to the CTA data acquisition.
	
The software system developed to control the devices of an SST-1M telescope is described, as well as the interface between the telescope abstraction to the CTA central control and the data acquisition system.}
	\FullConference{The 34th International Cosmic Ray Conference,\\
		30 July- 6 August, 2015\\
		The Hague, The Netherlands}
\begin{document}
	\section{Introduction}
		\input{intro.tex}

	\section{Software status for the device subsystems}\label{devices}
		\input{devices.tex}
	\section{Conclusions}
		\input{plans.tex}

	\appendix
	\section*{Acknowledgements}
		\input{acknowledgements.tex}
	\input{acronyms.tex}
	\bibliographystyle{JHEP}
	\bibliography{bibliography}
\end{document}

%% file: myaddontotemplates.tex
\usepackage[german,italian,french,english]{babel} 
\usepackage[T1]{fontenc}
\usepackage{textcomp}
\usepackage{cite,mcite}
\usepackage{amsmath,amstext,amssymb,mathtools,amsthm,slashed,cancel,turnstile,feynmf,wasysym}
\usepackage{longtable,multirow,tabularx,colortbl}
\usepackage{graphicx,subfigure,picture,epstopdf}
\usepackage{xspace,xcolor}
\usepackage{forarray,xparse,xkeyval,environ,ifthen}
\usepackage[capitalise]{cleveref}
\usepackage[printonlyused]{acronym}

\makeatletter
\newcommand*\rel@kern[1]{\kern#1\dimexpr\macc@kerna}
\newcommand*\widebar[1]{%
  \begingroup
  \def\mathaccent##1##2{%
    \rel@kern{1.2}
    \overline{\rel@kern{-1.1}\macc@nucleus\rel@kern{-0.1}}
    \rel@kern{0.1}
  }%
  \macc@depth\@ne
  \let\math@bgroup\@empty \let\math@egroup\macc@set@skewchar
  \mathsurround\z@ \frozen@everymath{\mathgroup\macc@group\relax}%
  \macc@set@skewchar\relax
  \let\mathaccentV\macc@nested@a
  \macc@nested@a\relax111{#1}%
  \endgroup
}
\makeatother

\input{physics.tex} 

%% file: physics.tex
\usepackage{xspace,feynmf}
\input{fontsandenvs.tex}
\input{math.tex}

\def\gcm{\ifmmode {\mathrm{g/cm}^2}\else
                   {g/cm$^2$}\fi}%
\def\EeV{\ifmmode {\mathrm{\ Ee\kern -0.1em V}}\else
                   \textrm{Ee\kern -0.1em V}\fi}%
\def\PeV{\ifmmode {\mathrm{\ Pe\kern -0.1em V}}\else
                   \textrm{Pe\kern -0.1em V}\fi}%
\def\TeV{\ifmmode {\mathrm{\ Te\kern -0.1em V}}\else
                   \textrm{Te\kern -0.1em V}\fi}%
\def\MeV{\ifmmode {\mathrm{\ Me\kern -0.1em V}}\else
                   \textrm{Me\kern -0.1em V}\fi}%
\def\GeV{\ifmmode {\mathrm{\ Ge\kern -0.1em V}}\else
                   \textrm{Ge\kern -0.1em V}\fi}%
\def\keV{\ifmmode {\mathrm{\ ke\kern -0.1em V}}\else
                   \textrm{ke\kern -0.1em V}\fi}%
\def\MeV{\ifmmode {\mathrm{\ Me\kern -0.1em V}}\else
                   \textrm{Me\kern -0.1em V}\fi}%
\def\eV{\ifmmode {\mathrm{\ e\kern -0.1em V}}\else
                   \textrm{e\kern -0.1em V}\fi}%

%% file: fontsandenvs.tex
\newcommand{\ben}{\begin{enumerate}}
\newcommand{\een}{\end{enumerate}}
\newcommand{\bi}{\begin{itemize}}
\newcommand{\ei}{\end{itemize}}
\newcommand{\bc}{\begin{center}}
\newcommand{\ec}{\end{center}}

%% file: math.tex
\usepackage{amsmath,amstext,amssymb,mathtools,amsthm,slashed,cancel,turnstile,upgreek} 
\usepackage{xparse,ifthen}

\DeclareDocumentCommand{\balign}{s}{\IfBooleanTF{#1}{\begin{align*}}{\begin{align}}}
\DeclareDocumentCommand{\ealign}{s}{\IfBooleanTF{#1}{\end{align*}}{\end{align}}}



%% file: authors_compact.tex
\author{\speaker{A.~Porcelli}$^{,e}$ for the  CTA Consortium and for the SST-1M sub-consortium: 
W.~Bilnik$^a$, J.~B{\l}ocki$^b$, L.~Bogacz$^c$, J.~Borkowski$^f$, T.~Bulik$^d$, F.~Cadoux$^e$, A.~Christov$^e$, M.~Cury{\l}o$^b$, D.~della~Volpe$^e$, M.~Dyrda$^b$, Y.~Favre$^e$, A.~Frankowski$^f$, {\L}.~Grudnik$^b$, M.~Grudzi\'nska$^d$, M.~Heller$^e$, B.~Id\'zkowski$^g$, M.~Jamrozy$^g$, M.~Janiak$^f$, J.~Kasperek$^a$, K.~Lalik$^a$, E.~Lyard$^h$, E.~Mach$^b$, D.~Mandat$^i$, A.~Marsza{\l}ek$^b$, J.~Micha{\l}owski$^b$, R.~Moderski$^f$, M.~Rameez$^e$, T.~Montaruli$^e$, A.~Neronov$^h$, J.~Niemiec$^b$, M.~Ostrowski$^g$, P.~Pa\'sko$^j$, M.~Pech$^i$, A.~Porcelli$^{e}$, E.~Prandini$^h$, P.~Rajda$^a$, E.~jr~Schioppa$^e$, P.~Schovanek$^i$, K.~Seweryn$^j$, K.~Skowron$^b$, V.~Sliusar$^k$, M.~Sowi\'nski$^b$, \L.~Stawarz$^g$, M.~Stodulska$^g$, M.~Stodulski$^b$, I.~Troyano~Pujadas$^e$, S.~Toscano$^{h,l}$, R.~Walter$^h$, M.~Wi\c{e}cek$^a$, A.~Zagda\'nski$^g$, K.~Zi\c{e}tara$^g$, P.~\`Zychowski$^b$\\
\llap{$^a$}{AGH University of Science and Technology, al.Mickiewicza 30, 30-059 Krak\'ow, Poland}\\
\llap{$^b$}{Instytut Fizyki J\c{a}drowej im.\ H.\ Niewodnicza\'nskiego Polskiej Akademii Nauk, ul.\ Radzikowskiego 152, 31-342 Krak\'ow, Poland}\\
\llap{$^c$}{Department of Information Technologies, Jagiellonian University, ul.\ Reymonta 4, 30-059 Krak\'ow, Poland}\\
\llap{$^d$}{Astronomical Observatory, University of Warsaw, al.\ Ujazdowskie 4, 00-478 Warsaw, Poland}\\
\llap{$^e$}{DPNC - Universit\'e de Gen\'eve, 24 Quai Ernest Ansermet, CH-1211 Gen\'eve, Switzerland}\\
\llap{$^f$}{Nicolaus Copernicus Astronomical Center, Polish Academy of Science, ul.\ Bartycka 18, 00-716 Warsaw, Poland}\\
\llap{$^g$}{Astronomical Observatory, Jagiellonian University, ul.\ Orla 171, 30-244 Krak\'ow, Poland}\\
\llap{$^h$}{ISDC,  Observatoire de Gen\'eve, Universit\'e de Gen\'eve, 16 Chemin de Ecogia, CH-1290 Versoix, Switzerland}\\
\llap{$^i$}{Institute of Physics of the Czech Academy of Sciences, 17.\ listopadu 50, Olomouc \& Na Slovance 2, Prague, Czech Republic}\\
\llap{$^j$}{Centrum Bada\'n Kosmicznych Polskiej Akademii Nauk, 18a Bartycka str., 00-716 Warsaw, Poland}\\
\llap{$^k$}{Astronomical Observatory, Taras Shevchenko National University of Kyiv, Observatorna str., 3, Kyiv, Ukraine}\\
\llap{$^l$}{also with Vrije Universiteit Brussels, Pleinlaan 2 1050 Brussels, Belgium}}

%% file: intro.tex
The \ac{CTA} observatory is conceived to be the largest observatory for the \ac{VHE} gamma rays, exploring energies from $\sim10 \GeV$ to few hundreds of \TeV. To span such energies, three different kind of telescopes are proposed: \ac{LST} to cover from $\sim10$ to $\sim100 \GeV$, \ac{MST} from $\sim100$ to $\sim1 \TeV$ and \ac{SST} to measure above few \TeV.

Among the \ac{SST} projects, the \ac{SST-1M} is a 4 m Davies-Cotton telescope, which comprises proven technology for the telescope structure, and innovative electronics and photosensors for the camera. Its design is meant to be simple, low-budget and easy-to-build industrially\cite{SST-1M}. The construction of the prototype is progressing rapidly and the first run is expected for September 2015. 

\begin{figure}[tbp]
	\centering
	\includegraphics[width = .7\textwidth, trim= 0mm 49mm 0mm 33mm, clip]{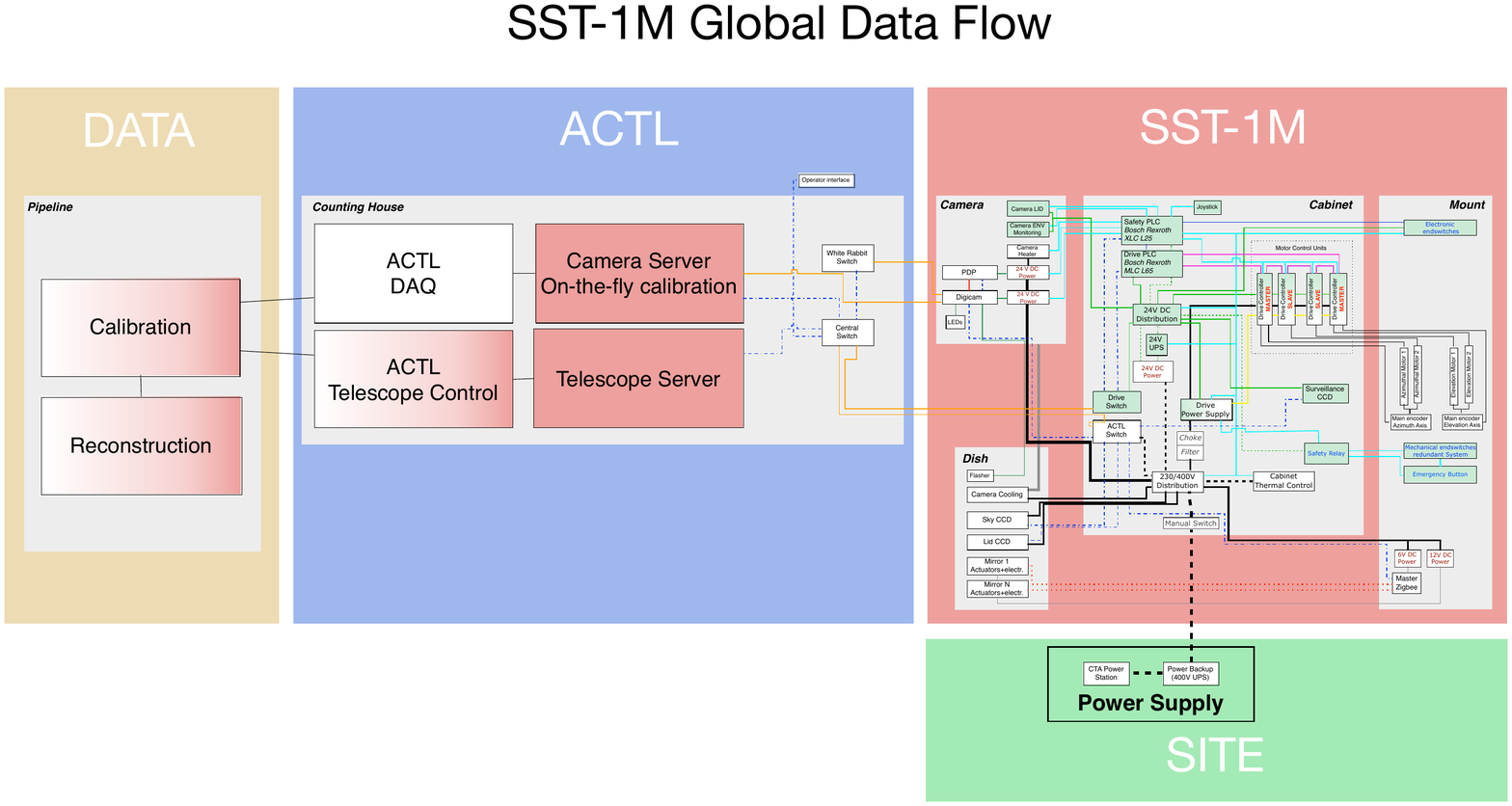}
	\caption[General view of the data flow between the \ac{SST-1M} and the \ac{ACTL}]{General view of the data flow between the \ac{SST-1M} and the \ac{ACTL}. The telescope abstraction (red block) communicates with the whole array via \acs{ACTL}, which will deal with the Data pipeline (yellow block). The scheme in the \ac{SST-1M} block (red) is not in the purpose of this paper and its full-size version is in\cite{SST-1M}. In the scheme, also the physically connection with the power supply on site is shown (green block).}
	\label{fig.genralview}
\end{figure}

In the meantime, the software to connect the control system of the telescope to the whole array of \ac{CTA} is under development. The whole array is controlled by the \ac{ACTL}\cite{ACTL}, based on the \ac{ACS} framework\cite{ACS1,ACS2}. It is the highest level of control and provides an abstract interface to the telescopes with configuration, logging and control services. The general view of the data flow between the \ac{SST-1M} abstraction and the \ac{ACTL} is shown in \Cref{fig.genralview}. The abstraction divides the telescope object (\ac{SST-1M} block in the figure) into five device subsystems, discussed in detail in \Cref{devices}: \ac{AMC}, \ac{CCD}, \ac{PLC}, \ac{CS} and \ac{CSC}.

The preferred policy for the \ac{ACTL} is to have a native \ac{OPC-UA} for each device. It is an industry standard communications protocol that uses an information model based on a full-mesh network of nodes. All the devices are visible through \ac{OPC-UA} servers, for which every device command and data point can be seen as nodes. Through a Java bridge, the nodes of the server are translated into object for the \ac{ACS} framework, which can be written in Java, C++ or Python languages. This framework enwraps subsystems, called components, to abstract the telescope as a single object. However a native \ac{OPC-UA} is not always available. In this case, it is possible to interface to a device directly via the \ac{ACS} component. 

The whole software is developed in a common operating system for the whole \ac{CTA} consortium: a virtual machine with a Scientific Linux 6.

%% file: devices.tex
\subsection{Active Mirror Control}\label{AMC}

\begin{figure}[tbp]
	\centering
	\includegraphics[width = .6\textwidth,trim= 34mm 59mm 8mm 88mm,clip]{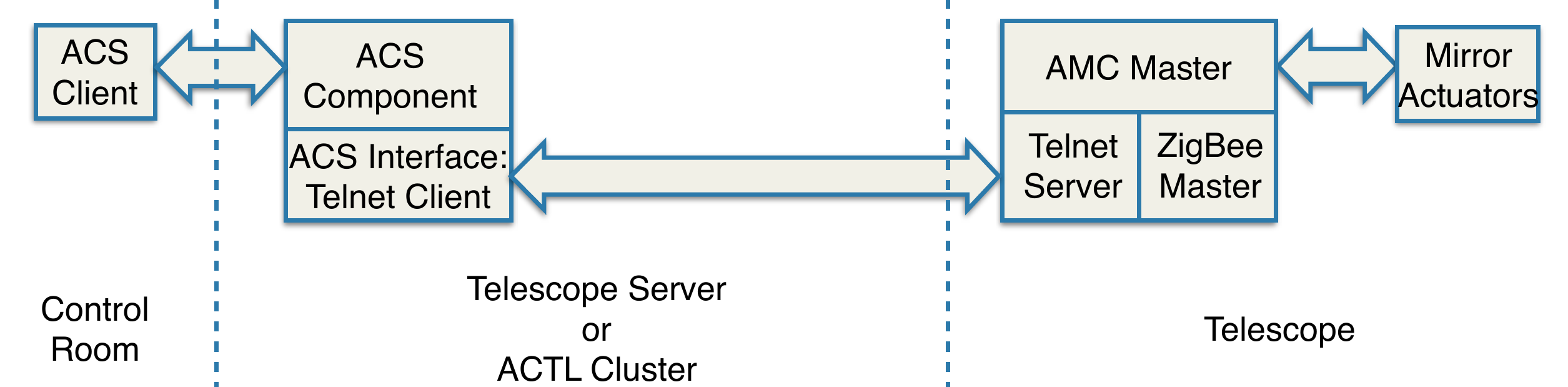}
	\caption{Scheme of the \acl{AMC} pipeline.}
	\label{fig.AMC}
\end{figure}

The \ac{AMC} is an interactive system to align the mirror facets mounted on the mirror dish\cite{SST-1M} with a precision of $1$ $\upmu$m. Each of the 18 facets has a set of two actuators for the $x$ and $y$ movement that can be moved with independent speeds specified by the user.

Each facet uses a telnet connection through a ZigBee master with a single IP address. No native \ac{OPC-UA} is provided, thus the interface to telnet is performed directly via \ac{ACS} components written in Java language. The pipeline is schematized in \Cref{fig.AMC}. The \ac{ACS}  component is completed and in beta-phase. Moreover, the software of the \ac{AMC} is used to test the skeleton of the web GUI, explained in \Cref{GUI}.

\subsection{CCD Camera}

\ac{SST-1M} has 3 different \acp{CCD}: ``Sky CCD'', ``LidCCD'' and ``Surveillance CCD''\cite{SST-1M}. The SkyCCD, used for the pointing, is mounted on a side of the dish support structure and provides an image of the sky with stars. The LidCCD, also mounted on the dish support structure but in the optical axis of the telescope, provides an image of the stars reflected on the camera lid and images of the pointing LEDs mounted in the camera focal plane. This camera is also used by the mirror alignment system. The Surveillance CCD is a surveillance camera.

All three \acp{CCD} are similar to the ones utilized by \ac{MST}, with the same software\cite{MST} adapted for the \ac{SST-1M} case. The software will control the image taking, image archiving and also slow control of the \acl{CCD} (power switching, temperature monitoring, error reporting). It also allows to identify the bright spots in the image (stars or LEDs images), the stars and their coordinates, and to calculate the required pointing calibration corrections. The \ac{ACS} component, written in Java language, communicates directly with the device, is fully functional and has been used continuously since almost two years in the \ac{MST} prototype. The adaptation to \ac{SST-1M} will be finalized by July 2015.

\subsection{Camera Server}\label{cameraserver}

\begin{figure}[tbp]
	\centering
	\includegraphics[width = .43\textwidth,trim= 0mm 8mm 0mm 34mm,clip]{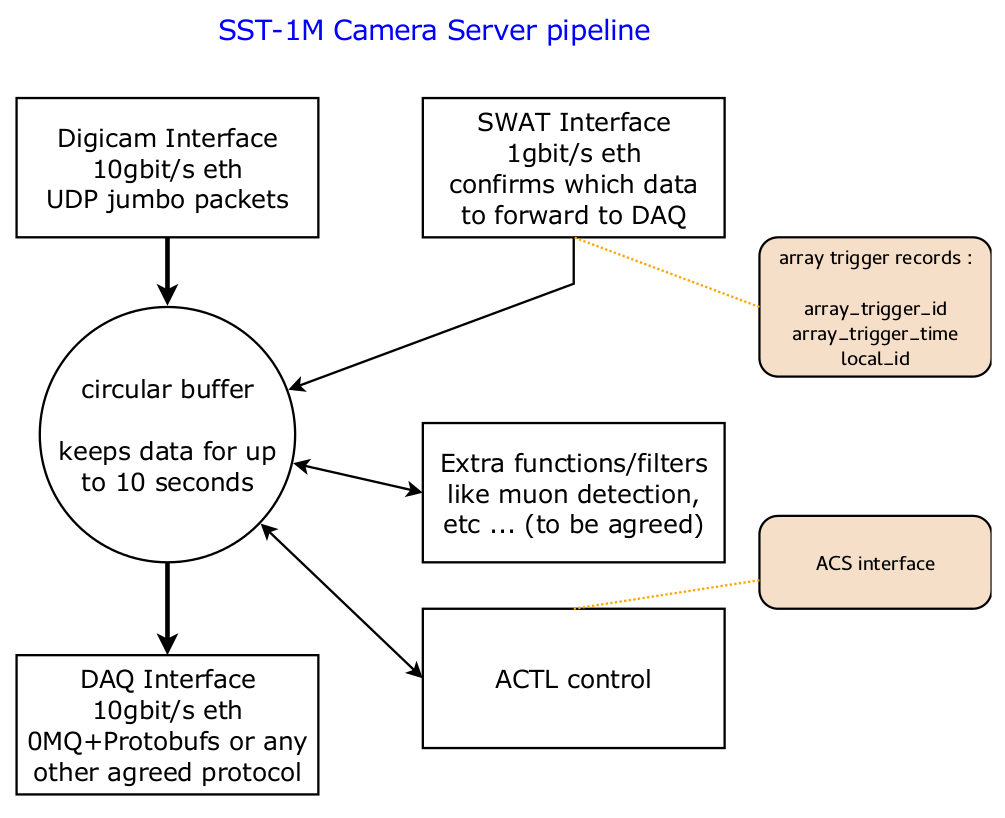}
	\caption{Scheme of the \acl{CS} pipeline. The colored boxes are the external interfaces to \ac{ACS} and to the record of the entire array.}
	\label{fig.CS}
\end{figure}

The \ac{CS} plays an important role in the bulk data reception\cite{SST-1M}. It consists of a Camera Server hardware and Camera Server software. It interfaces directly with the camera through DigiCam (via 10 Gbit/s link), the \ac{ACTL} and \ac{DAQ} (10 Gbit/s link), the \ac{CDTS} (1 Gbit/s link), the \ac{SWAT} (1 Gbit/s link). DigiCam is the camera Digital Trigger Readout System of \ac{SST-1M}\cite{SST-1M}. It controls the \ac{PDP}, as will be discussed in \Cref{CSC}. The \ac{SWAT} system checks which data are part of the triggered array and can be sent to the \ac{DAQ}\cite{ACTL}. \ac{CDTS} synchronizes and forwards the timestamp with a nanosecond precision, and provides an absolute clock with a $1~\upmu$s precision\cite{ACTL}.

When the bulk data packets are received, they are assembled into complete local camera triggered events and stored in memory for up to 10~s in a circular buffer. The local event  building verifies the proper sequencing of the packets, checks for data completeness, and filters out any duplicated, late, out-of-sync or erroneous packets. The timestamps are either embedded in the bulk data (if a native DigiCam \ac{WR}\footnote{The \acl{WR} is the common clock system of \ac{CTA}, which provides time-stamps with sub-nanosecond precision that are used to associate data to telescope events\cite{WRconference,WRjournal}.} interface is used), or read from \ac{CDTS} (if an UCTS board is used, as for current recommendation by the \ac{ACTL} developer group), and verified. Then \ac{CS} interfaces with the \ac{SWAT}, which replies on whether the local trigger is a member of an array trigger, part of any calibration or a special event. In the positive case, the data are forwarded to the \ac{DAQ} and the buffer is emptied. In the negative case, the stored information is discarded. In \Cref{fig.CS} such a scheme is shown. 

The \acl{CS} software tasks use the following communication protocols:
\bi
	\item for reception: the foreseen wire protocol is UDP jumbo packets; the exact contents of the packets are still to be finalized;
	\item for assembly: the DigiCam local camera triggered event size is about 64 Kbyte, which requires several UDP jumbo packets;
	\item interface between \ac{CS} and \ac{DAQ}: protocols and libraries recommended and approved by \ac{ACTL}; currently the proposed tools are ``0MQ library''\cite{0MQ} used for networking, and ``ProtoBufs''\cite{protobuf} for data encapsulation.
\ei

The \ac{SST-1M} Camera Server software will be integrated into \ac{ACTL} directly via the \ac{ACS} framework, using the C++ language. The software prototype to control the hardware is expected by July/August 2015 and its \ac{ACS} component by September 2015.

\subsection{Camera Slow Control}\label{CSC}

\begin{figure}[tbp]
	\centering
	\includegraphics[width = .6\textwidth,trim= 47mm 97mm 12mm 49mm,clip]{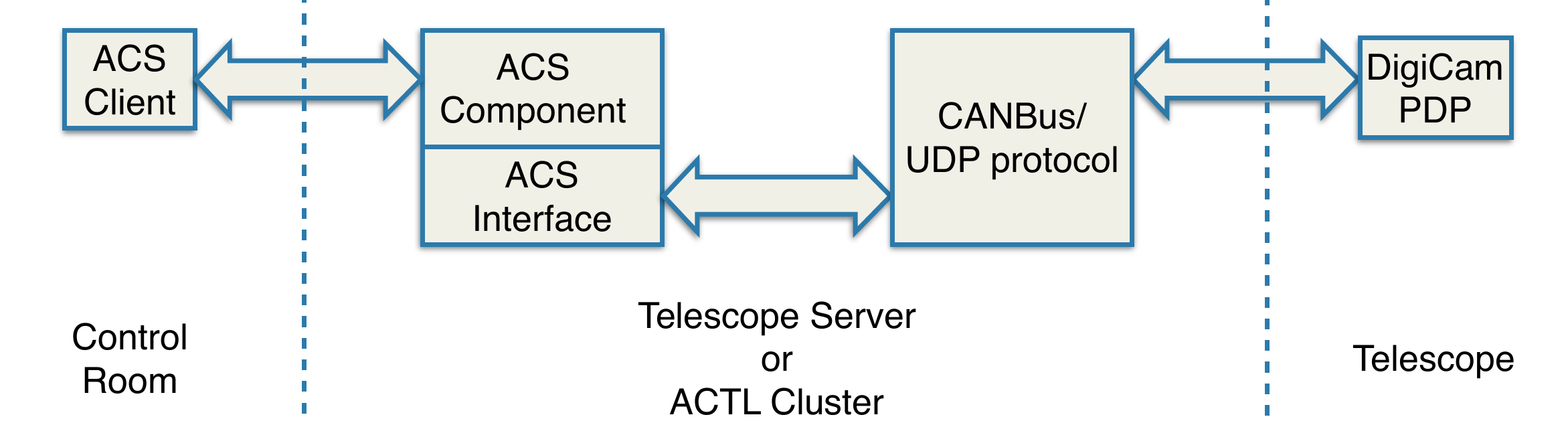}
	\caption{Scheme of the \acl{CSC} pipeline.}
	\label{fig.CSC}
\end{figure}

The \ac{CSC} manages the \ac{PDP} and is able to read the signals, temperature and set the bias Voltage for each single \ac{SiPM}\cite{SST-1M}. The read signal is sent and processed by DigiCam, which digitizes it through a \ac{FADC} with a sampling duration of 4~ns and applies the local camera trigger logic. If the event triggers the telescope camera, DigiCam will communicate with the \acl{CS} as explained in \Cref{cameraserver}. Moreover, since the operational point of the \acp{SiPM} depends on their temperature, the \ac{CSC} provides a compensation loop algorithm that reads the temperature and adjusts the bias Voltage accordingly.

The \ac{CSC} uses a CANBus integrated in the DigiCam, which communicates and sends commands to the \ac{PDP} via UDP connection. The \ac{CSC} does not have a native \ac{OPC-UA}, therefore the connection to the CANBus will be performed directly via \ac{ACS} components. Finally, the CANBus software drivers are provided with Python modules, thus the chosen language for \ac{ACS} is Python. Such a pipeline is schematized in \Cref{fig.CSC}.

The software development to control the \ac{CSC} is expected by mid July 2015. The development of the \ac{ACS} component is still in the early stages, but it will be finalized by September 2015.

\subsection{Programmable Logic Controller}

\begin{figure}[tbp]
	\centering
	\includegraphics[width = .7\textwidth]{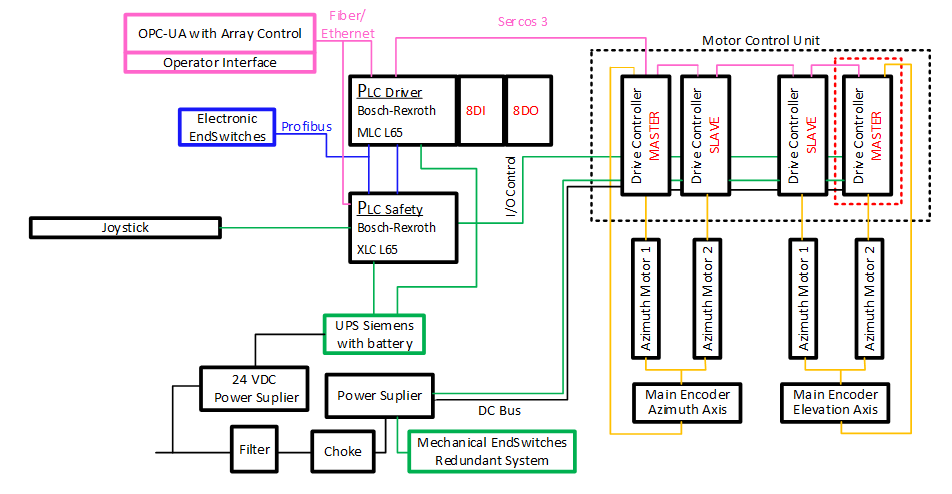}
	\caption{Scheme of the electrical and the communication interfaces of the \acl{PLC}.}
	\label{fig.PLC}
\end{figure}

\ac{SST-1M} has two different \acp{PLC}. The main one (also called ``drive'', \acs{dPLC}) is very similar to the \ac{MST} \ac{PLC} and uses the same software\cite{MST}. The auxiliary one is specific for this telescope is called \ac{sPLC}\cite{SST-1M}. It has embedded the Profibus and Sercos III communication interfaces. It supports signal modules as digital and analog inputs/outputs. The tasks of the \ac{sPLC} are to control the power supply of the main \ac{PLC} and  to guarantee a safe parking procedure that protects the camera from day light and mechanical damage.

In the case that the main \ac{PLC} fails or loses connection with the control room, the \ac{sPLC} will perform a safe parking. In the case of failure of all the encoders (embedded in the servo drives and mounted on the main axes), the Sercos III communication interface is used to support a second redundant parking procedure. This procedure uses the electrical limit switches mounted in appropriate limit positions. These switches are connected to the \ac{sPLC} with the Profibus gateway, which conducts the parking procedure according to the industrial homing method. Firstly, the homing of the azimuth axis is performed, driving the telescope to the azimuthal park position with some declared initial velocity. When the position is reached, the axis is moved again in the opposite direction with 10\% of the velocity for 2 seconds, and then moved back to the parking position once more with 1\% of the speed. In this way the accuracy of axis homing is $\sim0.01^\circ$. After the azimuth reaches the parking position, the same procedure is performed for the elevation, taking into account the protection of the camera from damages. Electrical and communication interfaces are shown in \Cref{fig.PLC}.

The \ac{sPLC} software development is in an advanced stage and expected to be completed by August 2015. The first tests were successfully done with a remote connection from Geneva\footnote{The telescope prototype is sited in the Instytut Fizyki J\c{a}drowej im.\ H.\ Niewodnicza\'nskiego Polskiej Akademii Nauk, Krak\'ow, Poland}. The \acp{PLC} are visible to the \ac{ACTL} via a native \ac{OPC-UA} server interface. The \ac{ACS} components to enwrap the \acp{OPC-UA} are in a development phase and are adapted from the \ac{MST} software written in Java language.

\subsection{GUI for the prototype}\label{GUI}

\begin{figure}[tbp]
	\centering
	\includegraphics[width = .7\textwidth,trim= 0mm 120 0mm 32mm,clip]{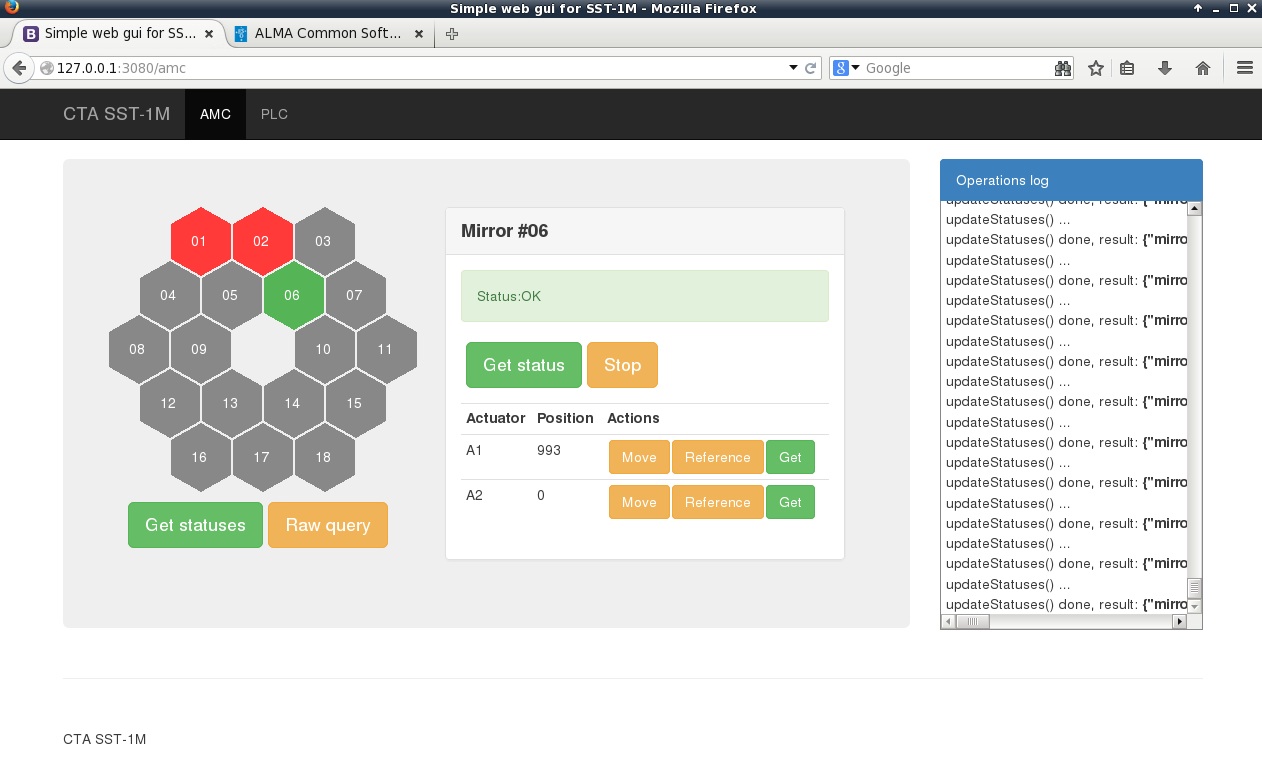}
	\caption{Screenshot of the GUI. The \ac{AMC} control is visualized.}
	\label{fig.GUI}
\end{figure}

To be able to control the telescope prototype, a web interface GUI is created. This interface will not be part of the \ac{ACTL}, but the sole purpose is to test the control of the subsystems and the telescope abstraction before they will be delivered to the \ac{ACTL} interface. The web interface is an HTML page that communicates with the \ac{ACS} components via a Java client. So far, only the \ac{AMC} (\Cref{AMC}) is dialoging with the GUI and is used to test it. A screenshot of the GUI is shown in \Cref{fig.GUI}.

%% file: plans.tex
\begin{figure}[tbp]
	\centering
	\includegraphics[width = .7\textwidth,trim= 0mm 52mm 0mm 25mm,clip]{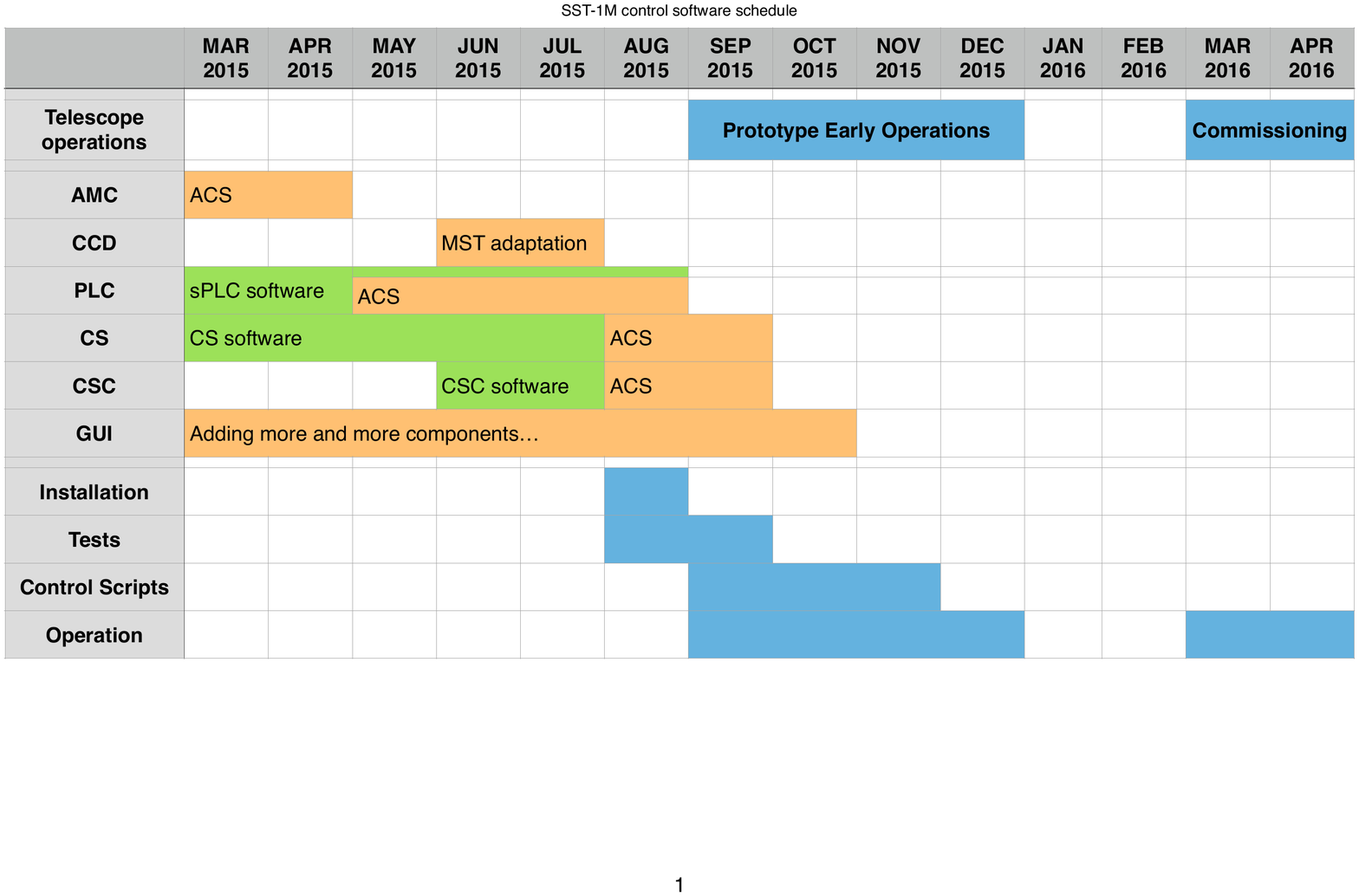}
	\caption{Planned schedule. In blue the general operations periods, in yellow the software development of the \ac{ACS} components for the \ac{ACTL} interfacing, and in green the software development of the control of the device low-level hardware.}
	\label{fig.schedule}
\end{figure}

The summary of the planned schedule described so far is shown in \Cref{fig.schedule}. The software development tasks are split into those for the \ac{ACS} components of the \ac{ACTL} interface (yellow areas) and the control of the device low-level hardware (green areas). In blue the planned operations for the telescope deployment are indicated to be compared with the software development schedule.

%% file: acknowledgements.tex
We gratefully acknowledge support from the agencies and organizations listed under Funding Agencies at this website: \href{http://www.cta-observatory.org/}{http://www.cta-observatory.org}.

In particular we are grateful for support from the NCN grant DEC-2011/01/M/ST9/01891 and the MNiSW grant Nr 498/1/FNiTP/FNiTP/2010 in Poland and the University of Geneva and the Swiss National Foundation in Switzerland.

%% file: acronyms.tex
	\begin{acronym}[OPC-UA]
		\acro{ACS}{ALMA Common Software}
		\acro{ACTL}{Central Array Control System}
		\acro{AMC}{Active Mirror Control}
		\acro{CCD}{CCD Camera}
		\acro{CDTS}{Clock Distribution and Trigger Timestamping}
		\acro{CS}{Camera Server}
		\acro{CSC}{Camera Slow Control}
		\acro{CTA}{Cherenkov Telescope Array}
		\acro{dPLC}{Drive Programmable Logic Controller}
		\acro{DAQ}{Data Acquisition System}
		\acro{FADC}{Fast Analog to Digital Converter}
		\acro{LST}{Large-Size Telescopes}
		\acro{MST}{Medium-Size Telescopes}
		\acro{OPC-UA}{Open Platform Communications Unified Architecture}
		\acro{PDP}{Photon Detection Plane}
		\acro{PLC}{Programmable Logic Controller}
		\acro{SiPM}{Silicon Photomultiplier}
		\acro{sPLC}{Safety Programmable Logic Controller}
		\acro{SST}{Small-Size Telescopes}
		\acro{SST-1M}{Small-Size Telescopes with single-mirror}
		\acro{SWAT}{Software Array Trigger}
		\acro{VHE}{Very High Energy}
		\acro{WR}{WhiteRabbit}
	\end{acronym}